\documentclass[useAMS,usenatbib]{mn2e}
\usepackage{graphicx}

\title[Colour Pairs for Constraining Stellar-Population Parameters]{Colour Pairs for Constraining the Age and Metallicity of Stellar Populations
\thanks{The data are available at the CDS or on request to the authors.}}
\author[Zhongmu Li and Zhanwen Han]{Zhongmu Li$^{1,2}$ \thanks{E-mail:
zhongmu.li@gmail.com} and Zhanwen Han$^{1}$\\
$^{1}$National Astronomical Observatories/Yunnan Observatory, the
Chinese Academy of Sciences, Kunming, 650011,
   China\\
$^{2}$Graduate School of the Chinese Academy of Sciences}
\begin{document}

\date{Accepted 1988 December 15. Received 1988 December 14; in original form 1988 October 11}

\pagerange{\pageref{firstpage}--\pageref{lastpage}} \pubyear{2002}

\maketitle

\label{firstpage}

\begin{abstract}
Using a widely used stellar population synthesis model, we study the
ability of using pairs of AB system colours to break the well-known
stellar age--metallicity degeneracy and give constraints on two
luminosity-weighted stellar-population parameters (age and
metallicity). The relative age and metallicity sensitivities of AB
system colours that relate to $u$, $B$, $g$, $V$, $r$, $R$, $i$,
$I$, $z$, $J$, $H$, and $K$ bands are presented, and the abilities
of various colour pairs for breaking the age--metallicity degeneracy
are quantified. Our results suggest that a few pairs of colours can
be used to constrain the two above stellar-population parameters.
This will be very useful for exploring the stellar populations of
distant galaxies. In detail, colour pairs [$(r-K), (u-R)$] and
[$(r-K), (u-r)$] are shown to be the best pairs for estimating the
luminosity-weighted stellar ages and metallicities of galaxies. They
can constrain two stellar-population parameters on average with age
uncertainties less than 3.89 Gyr and metallicity uncertainties less
than 0.34 dex for typical colour uncertainties. The typical age
uncertainties for young (Age $<$ 4.6 Gyr) populations and metal-rich
($Z$ $\geq$ 0.001) populations are small (about 2.26 Gyr) while
those for old populations (Age $\geq$ 4.6 Gyr) and metal-poor ($Z$
$<$ 0.001) populations are much larger (about 6.88 Gyr). However,
the metallicity uncertainties for metal-poor populations (about
0.0024) are much smaller than for other populations (about 0.015).
Some other colour pairs can possibly be used for constraining the
two parameters, too.

As a whole, the estimation of stellar-population parameters is
likely to reliable only for early-type galaxies with small colour
errors and globular clusters, because such objects contain less
dust. In fact, no galaxy is totally dust-free and early-type
galaxies are also likely have some dust, e.g., E(B-V)$\sim$ 0.05,
which can change the stellar ages about 2.5 Gyr and metallicities
($Z$) about 0.015. When we compare the photometric estimates with
previous spectroscopic estimates, we are shown some differences,
especially when comparing the stellar ages determined by two
methods. The differences mainly result from the young populations of
galaxies. Therefore, it is difficult to get the absolute values of
stellar ages and metallicities, but the results are useful for
getting some relative values.

In addition, our results suggest that colours relating to both
$UBVRIJHK$ and $ugriz$ magnitudes are much better than either
$UBVRIJHK$ colours or $ugriz$ colours for breaking the well-known
degeneracy. The results also show that the stellar ages and
metallicities of galaxies observed by the Sloan Digital Sky Survey
(SDSS) and the Two-Micron All-Sky Survey (2MASS) can be estimated
via photometry data.
\end{abstract}


\section{Introduction}

The formation and evolution of galaxies is one of the hottest topics
in astronomy and astrophysics. Great progress in the field has been
had (see, e.g., \citealt{Thomsen:1989}, \citealt{Kormendy:1989},
\citealt{Kodama:1998}, \citealt{vanDokkum:2003},
\citealt{Kauffmann:1998}, \citealt{Baugh:1996},
\citealt{Baugh:1998}, \citealt{Cole:2000}, \citealt{DeLucia:2006},
and \citealt{Thomas:1999}), and two luminosity-weighted
stellar-population parameters (age and metallicity) of galaxies are
crucial. A lot of stellar population synthesis models such as
\cite{Bruzual:2003}(hereafter BC03), \cite{Worthey:1994},
\cite{Vazdekis:1999} , \cite{Fioc:1997}, and \cite{Zhang:2005} have
been brought forward for stellar population studies and a great deal
of observational data have been supplied by big surveys such as the
Sloan Digital Sky Survey (SDSS) and the Two-Micron All-Sky Survey
(2MASS). However, it is difficult to measure the stellar ages and
metallicities of some distant galaxies (e.g., galaxies with
redshifts greater than 0.3) via the spectra-like methods. The reason
is that reliable spectra or line-strength indices (see, e.g.,
\citealt{Worthey:1994}, \citealt{Gallazzi:2005},
\citealt{Li:2006bpsstudy}) are usually available for some nearby
galaxies. If we can use one of important observational results,
i.e., photometry, we will be able to explore the stellar populations
of some distant galaxies. Because colours are easier to be obtained
than spectra and are independent of the distances of objects, they
are good candidates for studying the stellar populations of such
galaxies. Many works have been tried in this way, e.g.,
\cite{Dorman:2003}, \cite{Yi:2003}, \cite{Wu:2005},
\cite{James:2006}, \cite{Li:2007potential, Li:2007effects}, and
\cite{Kaviraj:2006better}. It seems possible to break the stellar
age--metallicity degeneracy \citep{Worthey:1994} by colours. Many
pairs of colours (hereafter colour pairs) are used in previous
works, e.g., $(U-R)$ and $(R-K)$ by \citet{Peletier:1996}, $(B-R)$
and $(R-K)$ by \citet{Bell:2000}, $(V-I)$ and $(V-K)$ by
\citet{Puzia:2002}, $(B-K)$ and $(J-K)$ by \citet{James:2006},
$(B-V)$ and $(B-K)$ by \citet{Li:2007potential}, $(B-V)$ and $(V-K)$
by \citet{Lee:2007}, and most colours used are only in $UBVRIJHK$
bands. As advocated by, e.g., \cite{Kong:2006}, colours relating to
both $UBVRIJHK$ and $ugriz$ bands, i.e., $(B-z)$ and $(z-Ks)$, can
be used to select galaxies with various redshifts. Similar colours
can possibly be used to break the stellar age--metallicity
degeneracy. We intend to give some investigations in this work,
using the BC03 stellar population synthesis model.

The paper is organized as follows. In Sect. 2, we briefly introduce
the BC03 model and the calculation of colours. In Sect. 3, we
present the age and metallicity sensitivities of colours. In Sect.
4, we try to search for colour pairs that can be used to constrain
stellar age and metallicity. In Sect. 5, we give our discussions and
conclusions.

\section{The BC03 model and calculation of colours}

The BC03 model is a widely used model in stellar population
synthesis study. Its standard model takes the \citet{Chabrier:2003}
initial mass function (IMF) and uses Padova 1994 library tracks to
calculate integrated colours. A few alternative stellar spectral
libraries are considered to give the spectral predictions of simple
stellar populations (SSPs). The model provides us magnitudes and
colours on both Johnson-Cousins-Glass ($UBVRIJHKLM$ bands) and AB
($ugriz$ bands) systems. More detailed information about the model
please refer to \citet{Bruzual:2003}. In this work, $BVRIJHK$
magnitudes on AB system are recalculated from those given on
Johnson-Cousins-Glass system, by taking -0.1, 0.0, 0.2, 0.45, 0.9,
1.4, and 1.9 as the differences between the zero points of AB system
and Johnson-Cousins-Glass system magnitudes, in $B$, $V$, $R$, $I$,
$J$, $H$ and $K$ bands, respectively \footnote{
http://www.astro.livjm.ac.uk/$^{\sim}$ikb/convert-units/node1.html}.

\section{age and metallicity sensitivities of colours}

We study the age and metallicity sensitivities by a relative
metallicity sensitivity ($rms$) technique, which was used by
\citet{Worthey:1994} and \citet{Li:2007potential}, and find age- and
metallicity-sensitive colours based on the $rms$ of colours. The
$rms$ method estimates the $rms$ of each colour by the ratio of
percentage change of age to that of metallicity when they lead to
the same change in a colour respectively. Colours with large $rms$
($>$1.0) are more sensitive to metallicity and those with small
$rms$ ($<$1.0) to stellar age. Following \citet{Li:2007potential},
we calculated the $rms$ of each colour in this work.

The $rms$ of 66 AB system colours are calculated in the work. The
detailed data are listed in Table 1.  As we see, $(B-K)$, $(R-K)$,
$(u-K)$, $(I-J)$, $(V-K)$, $(r-K)$, $(i-J)$, $(I-K)$, $(R-J)$,
 and $(g-J)$ are more sensitive to stellar metallicity while $(J-H)$,
$(i-I)$, $(R-H)$, $(z-K)$, $(V-H)$, $(u-R)$, $(B-V)$, $(B-g)$,
$(u-r)$, and $(g-H)$ to stellar age. We select the former ten
colours as age-sensitive colours while the latter ten as
metallicity-sensitive colours.

\begin{table}
 \caption{Relative metallicity sensitivities of 66 AB system colours.}
 \label{symbols}
 \begin{tabular}{llllll}
  \hline
   colour  &$rms$     &colour   &$rms$   &colour    &$rms$ \\
 \hline
$(B-K)$  &2.0294   &$(I-z)$  &0.8218   &$(B-H)$  &0.5758\\
$(R-K)$  &1.8610   &$(i-K)$  &0.8094   &$(u-g)$  &0.5654\\
$(u-K)$  &1.6904   &$(g-i)$  &0.7863   &$(g-V)$  &0.5400\\
$(I-J)$  &1.6125   &$(B-r)$  &0.7554   &$(i-H)$  &0.5178\\
$(V-K)$  &1.5774   &$(B-R)$  &0.7335   &$(H-K)$  &0.5105\\
$(r-K)$  &1.4114   &$(V-I)$  &0.7237   &$(R-I)$  &0.5074\\
$(i-J)$  &1.4031   &$(g-r)$  &0.7027   &$(r-R)$  &0.5012\\
$(I-K)$  &1.2654   &$(g-R)$  &0.6952   &$(r-H)$  &0.4834\\
$(R-J)$  &1.2160   &$(V-R)$  &0.6896   &$(J-K)$  &0.4707\\
$(g-J)$  &1.1094   &$(V-r)$  &0.6877   &$(R-i)$  &0.4675\\
$(B-z)$  &1.0914   &$(u-I)$  &0.6725   &$(u-V)$  &0.4440\\
$(u-J)$  &1.0474   &$(r-i)$  &0.6722   &$(z-H)$  &0.4180\\
$(B-J)$  &1.0455   &$(R-z)$  &0.6675   &$(g-H)$  &0.4116\\
$(V-J)$  &1.0145   &$(u-B)$  &0.6632   &$(u-r)$  &0.4000\\
$(V-z)$  &1.0069   &$(r-z)$  &0.6534   &$(B-g)$  &0.3984\\
$(u-z)$  &0.9985   &$(g-z)$  &0.6404   &$(B-V)$  &0.3908\\
$(z-J)$  &0.9805   &$(r-I)$  &0.6382   &$(u-R)$  &0.3887\\
$(g-K)$  &0.9719   &$(u-H)$  &0.6320   &$(V-H)$  &0.3835\\
$(r-J)$  &0.9247   &$(i-z)$  &0.6151   &$(z-K)$  &0.3637\\
$(B-i)$  &0.8788   &$(u-i)$  &0.5909   &$(R-H)$  &0.3344\\
$(V-i)$  &0.8719   &$(I-H)$  &0.5867   &$(i-I)$  &0.2262\\
$(B-I)$  &0.8414   &$(g-I)$  &0.5856   &$(J-H)$  &0.1670\\
  \hline
 \end{tabular}
 \end{table}

\section{colour pairs for breaking the age--metallicity degeneracy}

\subsection{Colour pairs for general studies}
Using the 10 metallicity-sensitive colours and the 10 age-sensitive
colours presented in the last paragraph of Sect. 3, we buildup 100
colour pairs. Each colour pair includes a metallicity-sensitive
colour and an age-sensitive colour. Using the colour pairs one by
one, we fit the stellar ages and metallicities of 500 testing
stellar populations. To make the results useful for estimating the
parameters of all kinds of stellar populations, the ages and
metallicities of the testing populations are generated randomly
within the ranges of 0.1--15 Gyr and 0.0001--0.05, respectively. The
averages of uncertainties in age and metallicity, $\overline{\Delta
t}$ and $\overline{\Delta Z}$, are then calculated by taking typical
uncertainties for input colours. The typical uncertainties for $U$,
$B$, $V$, $R$, $I$, $J$,  $H$,  $K$, $u$,  $g$,  $r$,  $i$, and $z$
magnitudes are taken as 0.109, 0.116, 0.059, 0.03, 0.07, 0.08, 0.09,
0.126, 0.11, 0.01, 0.007, 0.007 and 0.012 mag, respectively. These
values are estimated using the data supplied by the NASA/IPAC
Extragalactic Database (NED), the 2MASS and SDSS surveys. Then we
investigate the ability of each colour pair for breaking the
well-known stellar age--metallicity degeneracy by comparing the
$\overline{\Delta t}$ and $\overline{\Delta Z}$. A least-square
method is used to fit the ages and metallicities of stellar
populations in the work. The uncertainties in stellar-population
parameters are estimated by taking the maximum uncertainties in the
two parameters when considering the uncertainties of colours. One
can refer to \citet{Denicolo:2005} or \citet{Li:2006bpsstudy} for
more details. The main results are listed in Table 2.

Because colour pairs that can well break the stellar
age--metallicity degeneracy lead to small uncertainties in stellar
age and metallicity, pairs with small $\overline{\Delta t}$ and
$\overline{\Delta Z}$ are better for constraining two
stellar-population parameters. However, as we see, some colour pairs
have only small $\overline{\Delta t}$ or small $\overline{\Delta
Z}$. In this case, it is difficult to compare the abilities of
various colour pairs. We defined a parameter, uncertainty parameter
($UP$), to solve this problem. The $UP$ is calculated by taking the
average of the relative uncertainties of stellar ages and
metallicities of the 500 testing stellar populations. According to
the calculation of $UP$, colour pairs with small $UP$s are more
suitable for breaking the stellar age--metallicity degeneracy. The
$UP$s of colours are shown in Table 2, together with
$\overline{\Delta t}$s and $\overline{\Delta Z}$s. Considering
actual applications, we only list the results of colour pairs that
have $UP$s smaller than 2.0 when taking typical uncertainties for
colours. In the table, colours are sorted by an increasing order of
$UP$. Note that we also list the average metallicity uncertainty in
dex. As we see, [$(r-K), (u-R)$] and [$(r-K), (u-r)$] are the best
colour pairs for breaking the well-known degeneracy. Given typical
uncertainties for colours, the two colour pairs can constrain
stellar-population parameters with relative uncertainties smaller
than about 96\%, which corresponds to average uncertainties in
stellar age and metallicity smaller than 3.89\, Gyr and 0.34 dex,
respectively. Some other pairs, i.e., [$(R-K), (u-R)$], [$(I-K),
(u-R)$], [$(R-K)], (u-r)$] and [$(i-J), (u-R)$] can possibly be used
to give constraints on the stellar ages and metallicities of
galaxies as they have small $UP$s ($\leq$ 1.13) for typical
uncertainties of colours. In addition, colour pairs relating to both
$UBVRIIHJK$ bands and $ugriz$ bands are shown to be much better than
those only relating to one of the two kinds of bands. To see the
abilities of colour pairs for breaking the age--metallicity
degeneracy, we show the colour--colour grids of four pairs in Fig.
1.

However, as we note from Table 2, the uncertainties in two
stellar-population parameters are very big. This results from the
large observational uncertainties. Because the observational
uncertainties depend on surveys and they will possibly be decreased
in future surveys, we tried to find the best pairs for various
uncertainties (0.02, 0.05 and 0.10) in colours. We assumed that all
colours have the same uncertainty in each test. The results are
shown in Table 2. As we see, if the uncertainties of colours are
less than 0.05 mag, the stellar ages and metallicities of galaxies
can be constrained with uncertainty less than 1 Gyr and 0.0076,
respectively, via $(I-K)$ and $(u-R)$.

\subsection{Colour pairs for special studies}
Because we often need to estimate the ages and metallicities of some
special stellar populations, e.g., the old populations of globular
clusters, we tried to find some colour pairs that are suitable for
estimating the two luminosity-weighted stellar-population parameters
of such populations. In detail, the best colour pairs for estimating
the stellar ages and metallicities of young ($t$ $<$ 4.6 Gyr), old
($t$ $\geq$ 4.6 Gyr), metal poor ($Z$ $<$ 0.001), and metal rich
($Z$ $\geq$ 0.001) stellar populations are found by taking the above
typical uncertainties of colours. The best ten colour pairs for
studying each kind of special population are listed in Table 3. As
we see, some colour pairs for studying different kinds of stellar
populations are various. However, [$(r-K), (u-r)$] can be used to
constrain the age and metallicity of all kinds of stellar
populations. We can also find that when colour pair [$(r-K), (u-r)$]
is used to study the stellar-population parameters of various
populations, the uncertainties of results are different. The age
uncertainties of old or metal-poor populations are usually larger
than those of young or metal-rich populations.

\subsection{Composite colour pairs}
In practice, we can also use colour pairs including magnitudes on
different systems, as colours relating to the same bands but on
different systems usually have similar properties for breaking the
well-known stellar age--metallicity degeneracy \citep{Worthey:1994}.
For example, we can use colour pair [$(r-Ks), (u-r)$], in which $ur$
magnitudes are on AB-system and $Ks$ magnitude on 2MASS system
instead of [$(r-K), (u-r)$], in which all magnitudes are on
AB-system. Of course, we can use Johnson-Cousins-Glass system
colours together with AB system colours.

In the work, we analyzed colour pairs relating to five AB-system
bands ($ugriz$) and three 2MASS bands ($JHKs$). According to the
results, colour pairs [$(u-r),(r-Ks)$], [$(u-r), (i-J)$] and
[$(u-Ks), (z-Ks)$] are more suitable for constraining
stellar-population parameters than others. The $UP$s of the three
colour pairs can refer to Table 2. Because these pairs have
colour-colour grids similar to those shown in Fig. 1, we do not show
them here.

\begin{table*}
\begin{flushleft}
  \caption{The abilities of various colour pairs for breaking the age--metallicity degeneracy.
  Here $\overline{\Delta t}$ and $\overline{\Delta Z}$ (or $\overline{\Delta [Z/{\rm H}]}$) are the average uncertainties
  of stellar ages and metallicities, while $UP$ is the uncertainty parameter, i.e.,
  the average of relative uncertainties
  of ages and metallicities of 500 testing stellar populations.}
  \begin{tabular}{@{}lcccccccccccccc}
  \hline
  Uncertainty (mag)&&&typical&&\ &&0.02& &&0.05& &&0.10\\
  \hline
  Colour pair   &$\overline{\Delta t}$ &$\overline{\Delta Z}$ &$\overline{\Delta [Z/{\rm H}]}$ &$UP$ &\  &$\overline{\Delta t}$ &$\overline{\Delta Z}$ &$UP$  &$\overline{\Delta t}$ &$\overline{\Delta Z}$ &$UP$  &$\overline{\Delta t}$ &$\overline{\Delta Z}$ &$UP$\\
                &(Gyr)  &       & (dex) &       &\  &(Gyr)  &       &        & (Gyr) &       &        & (Gyr) &       &      \\
 \hline
$[(r-K),(u-R)]$ &  3.89 &0.0176 &  0.33 &  0.95 &\  &  0.37 &0.0047 &  0.17  &  1.01 &0.0085 &  0.35  &  2.13 &0.0145 &  0.66\\
$[(r-K),(u-r)]$ &  3.57 &0.0176 &  0.34 &  0.96 &\  &  0.40 &0.0045 &  0.17  &  1.14 &0.0093 &  0.40  &  2.58 &0.0146 &  0.78\\
$[(R-K),(u-R)]$ &  4.15 &0.0191 &  0.37 &  1.06 &\  &  0.33 &0.0044 &  0.16  &  0.96 &0.0084 &  0.35  &  1.98 &0.0142 &  0.63\\
$[(I-K),(u-R)]$ &  3.89 &0.0217 &  0.44 &  1.09 &\  &  0.27 &0.0032 &  0.12  &  0.89 &0.0076 &  0.31  &  2.33 &0.0131 &  0.58\\
$[(R-K),(u-r)]$ &  4.33 &0.0192 &  0.38 &  1.11 &\  &  0.32 &0.0040 &  0.15  &  1.04 &0.0087 &  0.37  &  2.19 &0.0142 &  0.67\\
$[(i-J),(u-R)]$ &  3.16 &0.0222 &  0.40 &  1.13 &\  &  0.45 &0.0078 &  0.25  &  1.28 &0.0148 &  0.55  &  3.54 &0.0228 &  1.19\\
$[(i-J),(u-r)]$ &  3.18 &0.0224 &  0.41 &  1.16 &\  &  0.50 &0.0089 &  0.29  &  1.44 &0.0148 &  0.63  &  3.71 &0.0206 &  1.19\\
$[(I-K),(u-r)]$ &  4.12 &0.0218 &  0.45 &  1.19 &\  &  0.28 &0.0034 &  0.12  &  0.92 &0.0078 &  0.32  &  2.34 &0.0133 &  0.61\\
$[(u-K),(z-K)]$ &  5.21 &0.0188 &  0.37 &  1.34 &\  &  0.34 &0.0041 &  0.16  &  1.48 &0.0081 &  0.39  &  4.01 &0.0144 &  0.90\\
$[(g-J),(z-K)]$ &  6.21 &0.0176 &  0.35 &  1.37 &\  &  0.48 &0.0073 &  0.31  &  2.27 &0.0082 &  0.44  &  4.56 &0.0132 &  0.98\\
$[(r-K),(R-H)]$ &  4.20 &0.0255 &  0.39 &  1.54 &\  &  0.48 &0.0073 &  0.31  &  3.54 &0.0181 &  1.08  &  3.79 &0.0211 &  1.26\\
$[(B-K),(z-K)]$ &  6.81 &0.0187 &  0.37 &  1.55 &\  &  0.32 &0.0072 &  0.28  &  2.13 &0.0084 &  0.43  &  4.61 &0.0146 &  0.96\\
$[(i-J),(g-H)]$ &  5.42 &0.0243 &  0.47 &  1.56 &\  &  0.42 &0.0096 &  0.34  &  2.61 &0.0168 &  0.77  &  7.28 &0.0257 &  2.26\\
$[(i-J),(z-K)]$ &  5.09 &0.0175 &  0.30 &  1.59 &\  &  0.42 &0.0096 &  0.34  &  2.47 &0.0158 &  0.83  &  5.65 &0.0146 &  1.71\\
$[(V-K),(i-I)]$ &  5.61 &0.0249 &  0.49 &  1.59 &\  &  0.42 &0.0096 &  0.34  &  2.80 &0.0201 &  1.05  &  3.08 &0.0203 &  1.10\\
$[(V-K),(u-R)]$ &  5.38 &0.0230 &  0.51 &  1.62 &\  &  0.38 &0.0050 &  0.18  &  1.19 &0.0101 &  0.44  &  2.82 &0.0152 &  0.80\\
$[(V-K),(u-r)]$ &  5.56 &0.0227 &  0.50 &  1.72 &\  &  0.57 &0.0052 &  0.21  &  1.32 &0.0107 &  0.48  &  3.13 &0.0151 &  0.98\\
$[(I-J),(z-K)]$ &  4.72 &0.0160 &  0.31 &  1.76 &\  &  2.43 &0.0110 &  0.93  &  2.72 &0.0110 &  0.74  &  5.54 &0.0153 &  2.05\\
$[(r-K),(g-H)]$ &  7.77 &0.0226 &  0.43 &  1.86 &\  &  0.66 &0.0064 &  0.29  &  2.99 &0.0096 &  0.55  &  6.85 &0.0176 &  1.44\\
$[(g-J),(u-r)]$ &  5.58 &0.0262 &  0.53 &  1.94 &\  &  0.91 &0.0140 &  0.53  &  2.36 &0.0182 &  1.02  &  6.45 &0.0264 &  2.74\\
$[(I-J),(J-H)]$ &  5.23 &0.0280 &  0.59 &  1.95 &\  &  0.91 &0.0140 &  0.53  &  2.36 &0.0182 &  1.02  &  6.06 &0.0215 &  2.06\\
$[(R-J),(u-r)]$ &  5.53 &0.0224 &  0.55 &  1.97 &\  &  0.40 &0.0074 &  0.23  &  1.95 &0.0178 &  0.94  &  5.09 &0.0233 &  1.88\\
$[(R-K),(g-H)]$ &  8.02 &0.0227 &  0.47 &  1.98 &\  &  0.60 &0.0077 &  0.35  &  2.79 &0.0094 &  0.54  &  6.13 &0.0167 &  1.27\\

\hline
\end{tabular}
\end{flushleft}
\end{table*}

\begin{figure*} 
  \includegraphics[angle=-90,width=176mm]{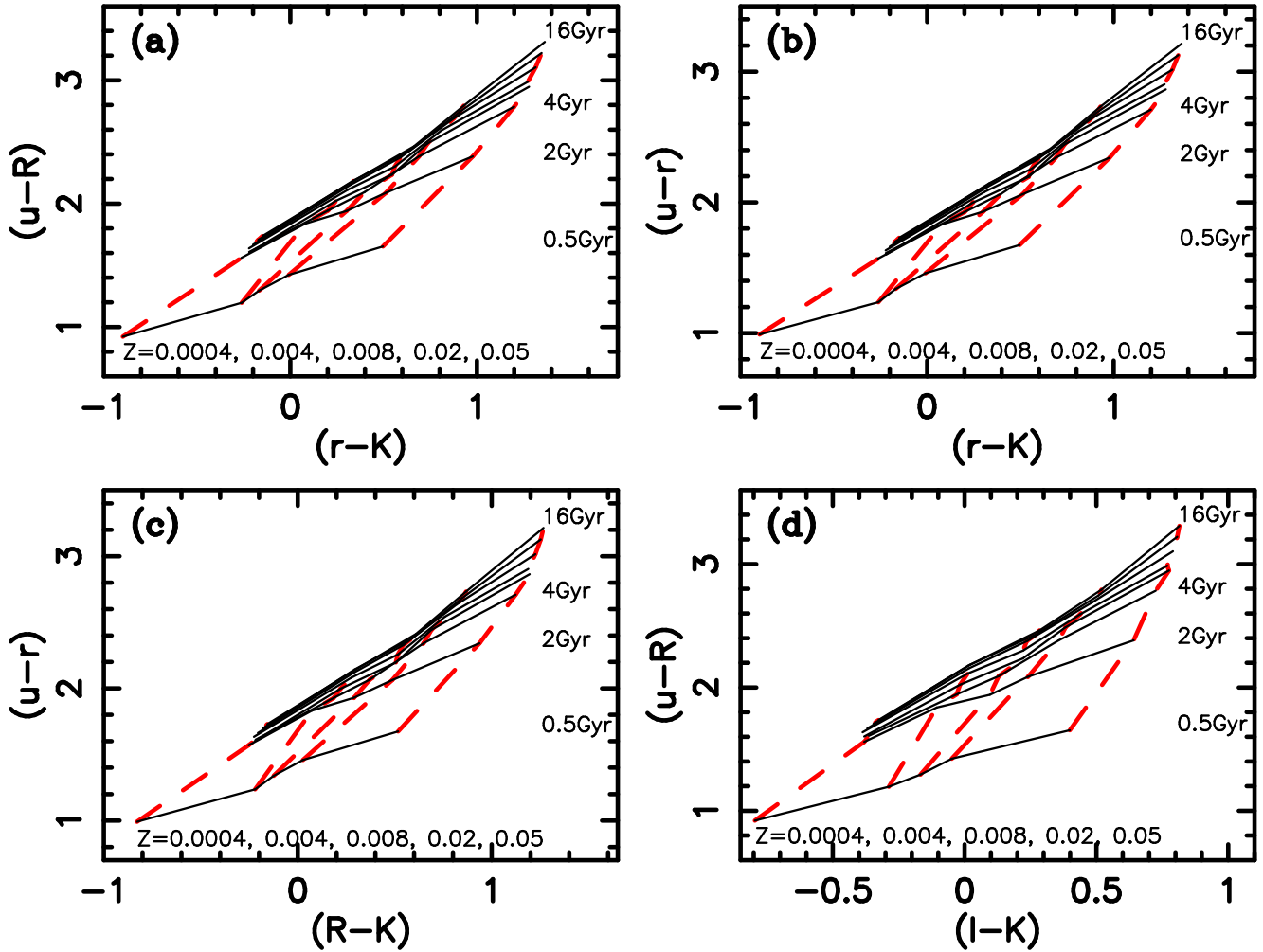}
  \caption{Colour--colour grids of four colour pairs that are suitable for constraining stellar age and metallicity.
 Colours are on AB system. Solid and dashed lines represent constant age and metallicity,
 respectively. Note that we did not mark for constant ages of 6, 8, 10,
 12 Gyr, as the limitation of the space.
 The four panels are for [$(r-K), (u-R)$], [$(r-K), (u-r)$], [$(R-K), (u-r)$], and [$(I-K), (u-R)$], respectively.}
\end{figure*}

\begin{table*}
\begin{flushleft}
  \caption{Best colour pairs for estimating the ages and metallicities of old (Age $\geq$ 4.6 Gyr), young (Age $<$ 4.6 Gyr), metal-rich ($Z \geq$ 0.001),
  and metal-poor ($Z <$ 0.001) stellar populations. Symbols have the same meanings as in Table 2.
  The values are calculated using the typical colour uncertainties.
  Note that the age uncertainties of metal-poor populations are not always right because some
  testing populations are out of the colour-colour grid when taking their colour uncertainties into account.}
  \begin{tabular}{@{}lrrrlrrrlrrr}
  \hline
  \hline
  \multicolumn{4}{c}{Old Population} &\multicolumn{4}{c}{Young or Metal-rich Population} &\multicolumn{4}{c}{Metal-poor Population}\\
  \hline
  Colour pair     &$\overline{\Delta t}$ &$\overline{\Delta Z}$ &$UP$&Colour pair &$\overline{\Delta t}$ &$\overline{\Delta Z}$ &$UP$&Colour pair &$\overline{\Delta t}$ &$\overline{\Delta Z}$ &$UP$\\
 \hline
$[(r-K),(u-r)]$   & 6.88    &0.0141    &0.73  &$[(r-K),(u-r)]$    &2.26   &0.0174    &0.76  &$[(r-K),(u-r)]$    &3.64   &0.0024    &4.96\\
$[(i-J),(u-r)]$   & 7.17    &0.0146    &0.73  &$[(i-J),(u-r)]$    &1.34   &0.0222    &0.78  &$[(r-K),(u-R)]$    &3.98   &0.0024    &4.96\\
$[(i-J),(u-R)]$   & 7.58    &0.0149    &0.75  &$[(i-J),(u-R)]$    &1.43   &0.0220    &0.81  &$[(R-K),(u-R)]$    &4.10   &0.0026    &5.52\\
$[(r-K),(u-R)]$   & 7.51    &0.0144    &0.76  &$[(r-K),(u-R)]$    &2.54   &0.0173    &0.81  &$[(g-J),(z-K)]$    &1.85   &0.0032    &5.69\\
$[(R-K),(u-r)]$   & 8.17    &0.0170    &0.89  &$[(R-K),(u-r)]$    &2.70   &0.0195    &0.85  &$[(R-K),(u-r)]$    &3.89   &0.0030    &5.79\\
$[(R-K),(u-R)]$   & 8.47    &0.0174    &0.91  &$[(R-K),(u-R)]$    &2.64   &0.0195    &0.86  &$[(B-K),(z-K)]$    &3.67   &0.0032    &6.18\\
$[(V-K),(u-R)]$   & 9.57    &0.0188    &1.07  &$[(R-J),(u-r)]$    &2.23   &0.0216    &0.93  &$[(i-J),(B-V)]$    &3.42   &0.0031    &6.23\\
$[(i-J),(i-I)]$   &11.29    &0.0177    &1.10  &$[(I-K),(u-R)]$    &2.46   &0.0235    &0.95  &$[(u-K),(z-K)]$    &6.33   &0.0029    &6.26\\
$[(V-K),(u-r)]$   & 9.19    &0.0187    &1.13  &$[(I-K),(u-r)]$    &2.50   &0.0235    &0.96  &$[(I-K),(B-V)]$    &3.87   &0.0030    &6.47\\
$[(I-K),(z-K)]$   & 9.79    &0.0170    &1.16  &$[(R-J),(u-R)]$
&2.29 &0.0250    &1.06  &$[(r-K),(B-V)]$    &3.44   &0.0031
&6.67\\
\hline
\end{tabular}
\end{flushleft}
\end{table*}

\section{Application of colours and colour pairs}

\subsection{Using colour pairs to constrain stellar-population parameters}
To test the application of colour pairs to estimate the ages and
metallicities of stellar populations, we select 1\,646 luminous
(absolute magnitude $M_{\rm r} < -22$ and $r$-band Petrosian
magnitude $<$ 17.77) early-type (concentration index $C \geq$ 2.8)
galaxies observed by both 2MASS and the second release of SDSS
(SDSS-DR2). All the sample galaxies have small magnitude
uncertainties ($<$ 0.15 mag). Note that only the galaxies with
colour-fitted stellar ages smaller than 15 Gyr and stellar
metallicities richer than 0.008 are selected for our sample
galaxies, because the age of the universe was shown to be smaller
than about 15 Gyr (e.g., \citealt{Shafieloo:2006}) and the results
for populations with metallicities poorer than 0.008 seems
unreliable as large uncertainties. Then we use $(r-K)$ and $(u-r)$
colours to estimate two stellar-population parameters of these
galaxies, according to the results shown in Tables 2 and 3. The
$K$-band magnitudes of galaxies are calculated from $Ks$-band
magnitudes supplied by 2MASS, by using the same method as
\cite{Bessell:2005}. The $k$-corrections of $Ks$-band magnitudes are
estimated as ${\rm-6log}(1+z)$, as used by \cite{Girardi:2003},
where $z$ is the redshift. The galactic extinctions of $K$-band
magnitudes are calculated using the model of \citet{Burstein:1982}.
The $u$ and $z$ bands magnitudes, and their $k$-corrections and
galactic extinctions are taken from SDSS. The differences between
SDSS magnitudes and AB magnitudes are taken into account according
to the values supplied by SDSS. In Fig. 2, the sample galaxies are
shown on the $(u-r)$ versus $(r-K)$ grid. For clearly, we only
plotted the error bars for the first 10 galaxies of our sample,
which are marked in black. We see that a rough estimation for the
two parameters of galaxies can be obtained, although the
uncertainties are somewhat big. As examples, we list the
stellar-population parameters of a few galaxies in Table 4, in which
the data of three subsets of galaxies are listed.
\begin{table*}
\begin{flushleft}
 \caption{Stellar ages and metallicities of old (Age $\geq$ 4.6 Gyr),
 young metal-rich (Age $<$ 4.6 Gyr and $Z \geq$ 0.02), and
 young metal-poor (Age $<$ 4.6 Gyr and $Z <$ 0.02) galaxies.
 The results are measured by two kinds of methods (photometric and spectroscopic).
 The photometric results are determined via $(u-K)$ and $(z-K)$ colours in the work, and the spectroscopic results are obtained by
 \citet{Gallazzi:2005}.
 The symbol ``objID'' is the unique SDSS identifier of each galaxy.
 Note that the definition of metal-poor galaxy here is different from that in other places of the paper,
 as the limitation of our sample galaxies.}
 \label{symbols}
 \begin{tabular}{ccclcllclcl}
  \hline
  \hline
  &&\multicolumn{4}{c}{Photometric results} &&\multicolumn{4}{c}{Spectroscopic results}
  \\
  \hline
  Population type&  objID &   Age&   Error&    $Z$&    Error &&   Age&  Error&    $Z$&    Error\\
      &        &  (Gyr)&  (Gyr)&       &          &&  (Gyr)& (Gyr)\\
 \hline
   &587727177921921120  &4.6 &$_{-1.5}^{+3.6}$ &0.0293 &$_{-0.0125}^{+0.0110}$ &&7.4 &$_{-2.7}^{+5.4}$ &0.0411 &$_{-0.0171}^{+0.0077}$\vspace{2mm}\\
Old  &587727225693143113  &5.7 &$_{-2.9}^{+10.7}$ &0.0300 &$_{-0.0176}^{+0.0143}$ &&6.8 &$_{-3.3}^{+3.7}$ &0.0291 &$_{-0.0088}^{+0.0207}$\vspace{2mm}\\
   & 587727225692160083  &9.1 &$_{-4.5}^{+5.4}$ &0.0416 &$_{-0.0108}^{+0.0082}$ &&6.7 &$_{-2.6}^{+2.9}$ &0.0350 &$_{-0.0127}^{+0.0141}$\vspace{2mm}\\
\hline
   &587727177921986670  &2.7 &$_{-0.4}^{+0.6}$ &0.0478 &$_{-0.0121}^{+0.0022}$ &&5.3 &$_{-2.8}^{+3.6}$ &0.0241 &$_{-0.0133}^{+0.0220}$\vspace{2mm}\\
Young and metal-rich&587727225160466548  &2.2 &$_{-0.7}^{+0.6}$ &0.0385 &$_{-0.0081}^{+0.0089}$ &&6.2 &$_{-3.0}^{+2.8}$ &0.0301 &$_{-0.0114}^{+0.0162}$\vspace{2mm}\\
   &587727225689538771  &2.7 &$_{-0.4}^{+0.5}$ &0.0432 &$_{-0.0124}^{+0.0068}$ &&6.0 &$_{-2.3}^{+3.0}$ &0.0305 &$_{-0.0109}^{+0.0146}$\vspace{2mm}\\
\hline
   &587727230524063879  &3.7 &$_{-1.0}^{+8.0}$ &0.0189 &$_{-0.0121}^{+0.0128}$ &&7.4 &$_{-3.6}^{+2.9}$ &0.0226 &$_{-0.0097}^{+0.0184}$\vspace{2mm}\\
Young and metal-poor&587727227302707220  &3.8 &$_{-0.9}^{+4.1}$ &0.0166 &$_{-0.0085}^{+0.0038}$ &&6.2 &$_{-2.6}^{+3.1}$ &0.0251 &$_{-0.0073}^{+0.0179}$\vspace{2mm}\\
   &588848898845114541  &2.9 &$_{-0.9}^{+2.7}$ &0.0192 &$_{-0.0044}^{+0.0097}$ &&3.5 &$_{-2.2}^{+2.5}$ &0.0135 &$_{-0.0054}^{+0.0304}$\\
\hline
 \end{tabular}
 \end{flushleft}
 \end{table*}
In fact, even if we take Lick indices for such work, the
uncertainties in stellar ages and metallicities of these galaxies
are very big because of the large observational uncertainties. For
our sample galaxies, the uncertainties in H$\beta$ and [MgFe]
indices are typically about 0.3 $\rm \AA$, which will lead to large
uncertainties in stellar-population parameters. The Lick indices and
uncertainties of our sample galaxies are taken from the Garching
SDSS catalogs (http://www.mpa-garching.mpg.de/SDSS/DR4/).
Furthermore, we found that 24 galaxies are out of the theoretical
grid of BC03 models. This possibly results from the effects of young
stellar populations of galaxies, the limitation of theoretical
models, and large observational uncertainties of colours. In
special, we found that most galaxies can fall inside the theoretical
colour-colour grid when taking their colour uncertainties into
account. This means that most of the 24 galaxies could have physical
ages, but as pointed out by \citet{Li:2007potential}, the presence
of young populations in such early-type galaxies can also make
metal-rich populations outside the theoretical colour-colour grids.

In the work, we fitted the stellar ages and metallicities of our
sample galaxies via BC03 SSPs with ages within 0.1--19.96 Gyr and
metallicities within 0.0001--0.05. In Fig. 3, we compare the
parameters determined by colours with those by a few Lick indices
\citep{Gallazzi:2005}. The results of \citet{Gallazzi:2005} were
obtained by comparing D4000, H$\beta$, H$\delta_A$+H$\gamma_A$,
[Mg$_{\rm 2}$Fe], and [MgFe]$'$ indices of galaxies to the values of
theoretical stellar populations and have taken the effects of young
populations (YSPs) into account. The detailed data about the sample
galaxies can be obtained on request to the authors or via the CDS in
the future. From panels (a) and (c) of Fig. 3, we find that the
stellar ages and metallicities determined by colours are
respectively smaller and somewhat richer than those determined by
Lick indices. The reason is that there are composite stellar
populations (CSPs) in galaxies and the YSPs make the results derived
from colours bias younger and richer in metal compared to those of
the dominant populations (DSPs) of galaxies \citep{Li:2007effects}.
In detail, because the age of a YSP affects the colours of a star
system more stronger than the mass fraction of the YSP, YSPs with
only a few percent stellar mass can make the colour-fitted stellar
populations younger and metal-richer than DSPs of galaxies if the
YSPs are not too old. Then in panels (b) and (d) Fig. 3, we compare
the stellar-population parameters measured by colours and corrected
for the effects of YSPs to those determined by Lick indices. The
correction is accomplished using the possible distributions of the
differences between the stellar-population parameters of the DSPs of
CSPs and the parameters derived from two colours of CSPs
\citep{Li:2007effects}. The above distributions were obtained by a
statistical method, in which the fractions of YSPs were assumed to
depend on the ages of the DSPs and YSPs, and the fraction of a YSP
was assumed to decline exponentially with decreasing age of the YSP
(see, e.g., \citealt{Thomas:2005}). In special, the distributions
can be used to give a rough correction for the effects of YSPs.
Here, we get the corrected stellar age of a galaxy by submitting
from the colour-fitted result a random value that fits to the
distribution of the difference between the stellar ages of DSPs and
those derived from two colours. A similar method is used to get the
corrected stellar metallicities. For more clearly, we show the
distributions of the uncorrected and corrected parameters in Fig. 4.
We are shown that after the correction, the distributions of stellar
ages obtained by two methods, especially the peaks, become more
similar after the correction (Fig. 4). However, it is also shown
that the distributions of stellar metallicities obtained by various
methods are similar. Therefore, it seems that without any
correction, we can get correct distributions of stellar
metallicities of luminous early-type galaxies via colours. Note that
because we used a least-square fitting method similar to
\citet{Denicolo:2005} and \citet{Li:2006bpsstudy}, the effects of
the uncertainties of colours were not taken into account here. The
uncertainties may effect the above distributions slightly.
Furthermore, the dust may contribute to the difference between the
two kinds of results. In fact, a small amount of dust can make
$(u-r)$ colours of galaxies redder, and then lead to additional
uncertainties in stellar ages. In detail, a dust of E($B-V$) = 0.05
will change the $(u-r)$ and $(r-K)$ colours about 0.12 mag, and then
lead to about 2.5 Gyr age uncertainty and 0.015 metallicity
uncertainty. In addition, the average stellar-population parameters
obtained by photometry and spectroscopy are found to be similar
(about 7 Gyr for age and 0.02 for metallicity). Therefore, although
it is difficult to get the accurate stellar-population parameters of
each galaxy via colours, we can get reliable values for the average
age and metallicity of a sample of galaxies, as can the
distributions of the two parameters.
\begin{figure*} 
\includegraphics[angle=-90,width=176mm]{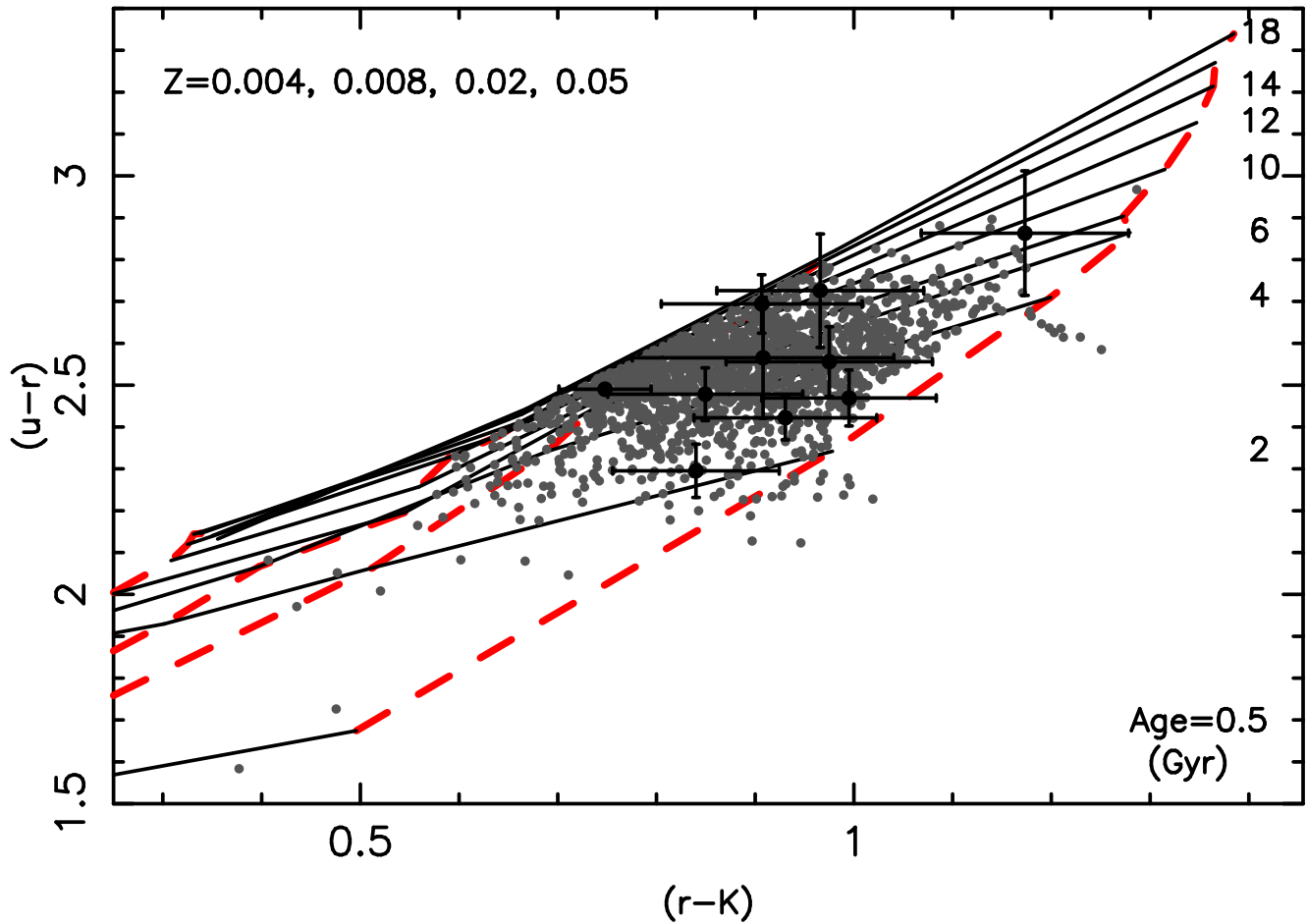}
 \caption{Our sample galaxies (1 646 galaxies) in the $(u-r)$ versus $(r-K)$
 grid. Dark points with error bars show the first 10 galaxies of our
 sample. The colour uncertainties of other galaxies can refer to those shown. Lines
 have the same meanings as in Fig. 1.
 Here we did not mark for the constant age of 8 and 16 Gyr.}
\end{figure*}

\begin{figure*} 
\includegraphics[angle=-90,width=176mm]{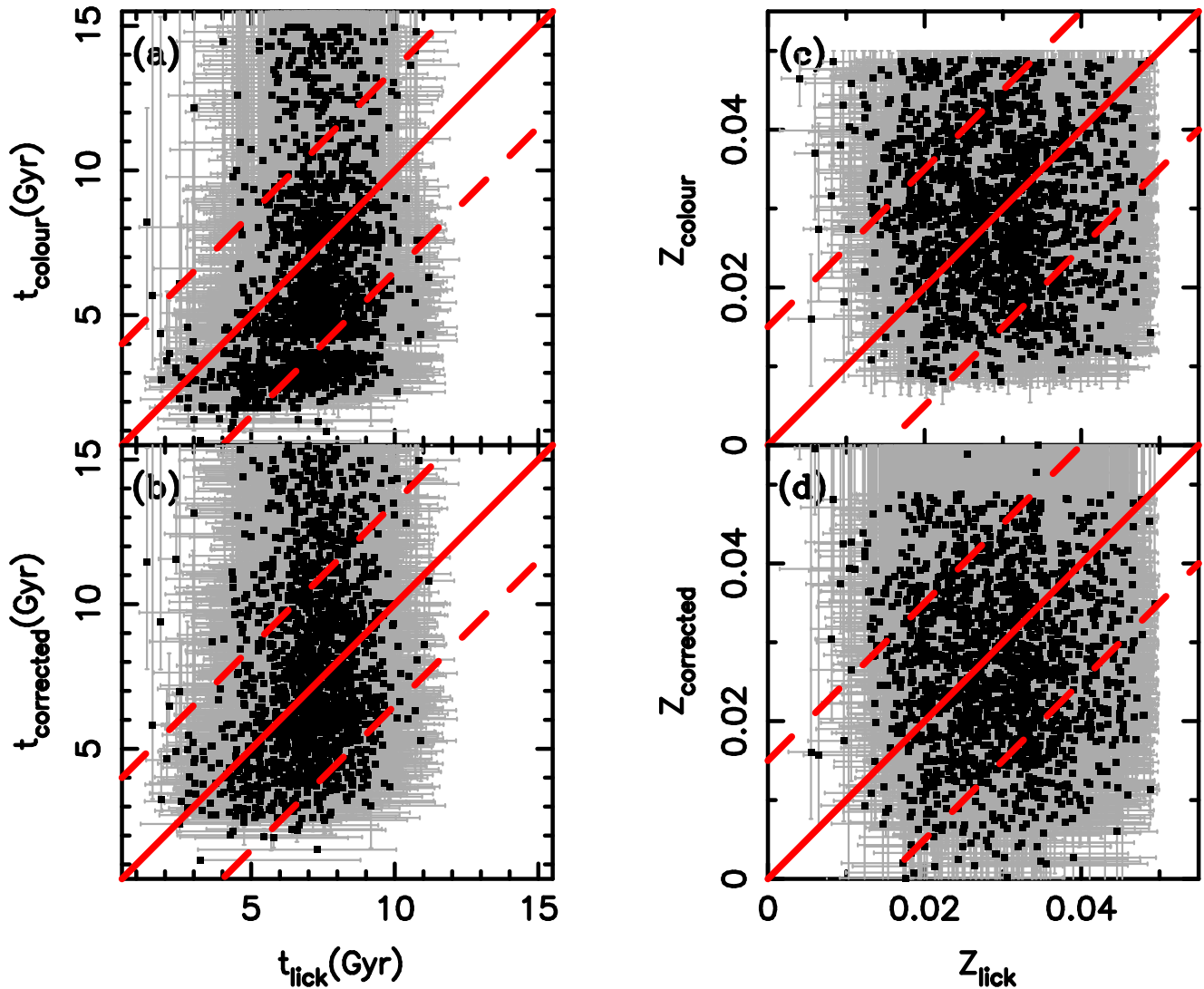}
 \caption{Comparison of stellar ages and metallicities derived from colours with those from Lick indices.
 The results derived from colours are obtained in this work and those
 derived from Lick indices were obtained by \citet{Gallazzi:2005}.
 The typical errors of stellar ages obtained by colours and Lick
 indices are about 3.80 Gyr and 4.17 Gyr, respectively.
 The symbol ``lick'' denotes the results of \citet{Gallazzi:2005}
 while ``colour'' means the results derived from $(r-K)$ and $(u-r)$.
 The suffix ``corrected'' means the results corrected for the effects of young stellar populations.
 Dashed lines show a $\pm$ 3.5 Gyr spread about the unity (solid) line for stellar ages in panels
 a) and b)
 while show a $\pm$ 0.015 spread in stellar metallicities in panels c) and d). }
\end{figure*}

\begin{figure} 
\includegraphics[angle=-90,width=88mm]{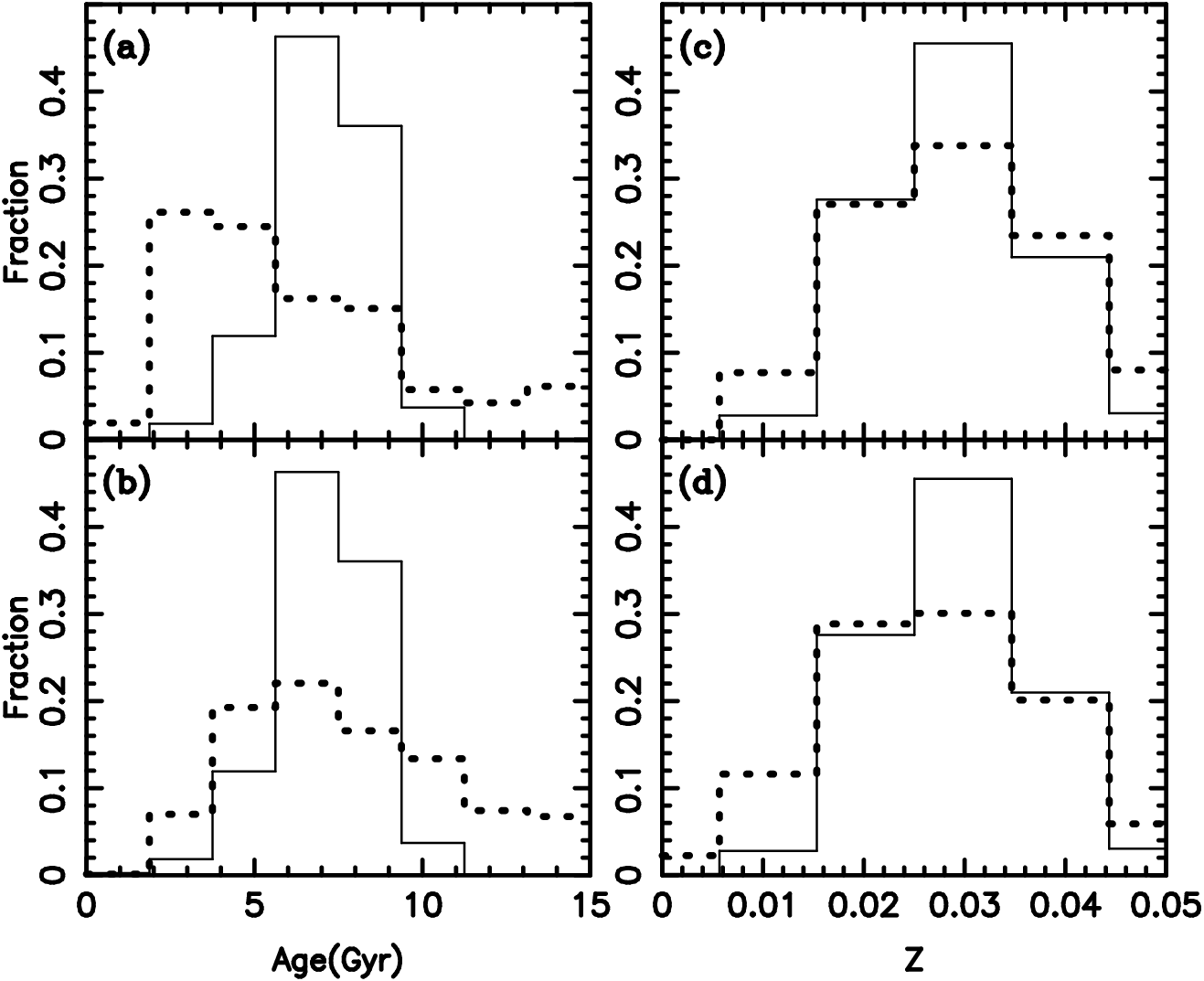}
 \caption{Comparison of the distributions of stellar-population parameters obtained by two kinds of methods.
 Solid lines are for the results obtained by Lick indices (\citealt{Gallazzi:2005}).
 Dotted lines in panels a) and c) show the results obtained by [$(r-K)$, $(u-r)$] and without any correction,
 while those in in panels b) and d) show the results corrected for the effects of young populations
 using the results of \citet{Li:2007effects}.}
\end{figure}

\subsection{Using colours in conjunction with spectroscopic indices}

Because some colours are shown to be sensitive to stellar age or
metallicity, we can possibly use colours in conjunction with Lick
indices to study stellar-population parameters. For example, we can
use a metallicity-sensitive colour together with an age-sensitive
Lick index to estimate the stellar ages and metallicities of
galaxies. We have a try in this work. The above galaxy sample are
used here. The stellar-population parameters are fitted via H$\beta$
and $(r-K)$ indices using the same method as Sect 5.1. The results
are compared to those determined by \citet{Gallazzi:2005} in panels
(a) and (b) of Fig. 5. Note that the results did NOT correct for the
effects of young populations. We see that using H$\beta$ instead of
$(u-r)$, the stellar ages obtained are closer to those of
\citet{Gallazzi:2005}. This means that line indices are affected by
young populations of galaxies more slightly. However, the fitted
stellar metallicities of some galaxies are poorer than those
obtained by \citet{Gallazzi:2005}. It should mainly result from the
effects of YSPs in galaxies. In fact, YSPs make the $(r-K)$ colour
of galaxies obviously bluer than those of the DSPs, but they affect
the H$\beta$ index more slightly. When $(r-K)$ is used in
conjunction with H$\beta$ to estimate the stellar-population
parameters of galaxies, the two indices will lead to more poor
metallicities compared those determined by a colour pair. One can
see a H$\beta$ versus [MgFe] grid for comparison, as the shape of
H$\beta$ versus $(u-r)$ grid is similar to that of H$\beta$ versus
[MgFe]. Therefore, it is difficult to use photometry in conjunction
with spectroscopy to estimate the metallicities of galaxies.
Furthermore, we tried to give some final results for the
stellar-population parameters of our sample galaxies using H$\beta$,
[MgFe], $(u-r)$ and $(r-K)$ indices together. Because the
uncertainties of colours and Lick indices are different, we use a
$\chi^{\rm 2}$ fit (see, e.g., \citealt{Press:1992}) to estimate the
stellar-population parameters. The stellar metallicities are shown
to be different significantly from those determined by previous
work. The comparison can be seen in panels (c) and (d) of Fig. 5.
The results did not correct for the effects of young populations,
either. Because the results of \citep{Gallazzi:2005} have taken the
effects of YSPs into account, the above results are certainly
affected by the YSPs in galaxies. The results may also effected by
the large observational uncertainties. However, it seems that the
YSPs effect the results stronger, because when we tried to estimate
the stellar-population parameters via H$\beta$ and [MgFe], we got
poorer metallicities than those of the previous work. As a whole,
our results suggest that colour pairs can be used more conveniently
for estimating the stellar-population parameters of distant
galaxies, as colours can be obtained more easily than spectra and
can constrain the stellar-population parameters with uncertainties
similar to Lick indices (typically 4 Gyr for age and 0.015 for
metallicity).

\begin{figure*} 
\includegraphics[angle=-90,width=176mm]{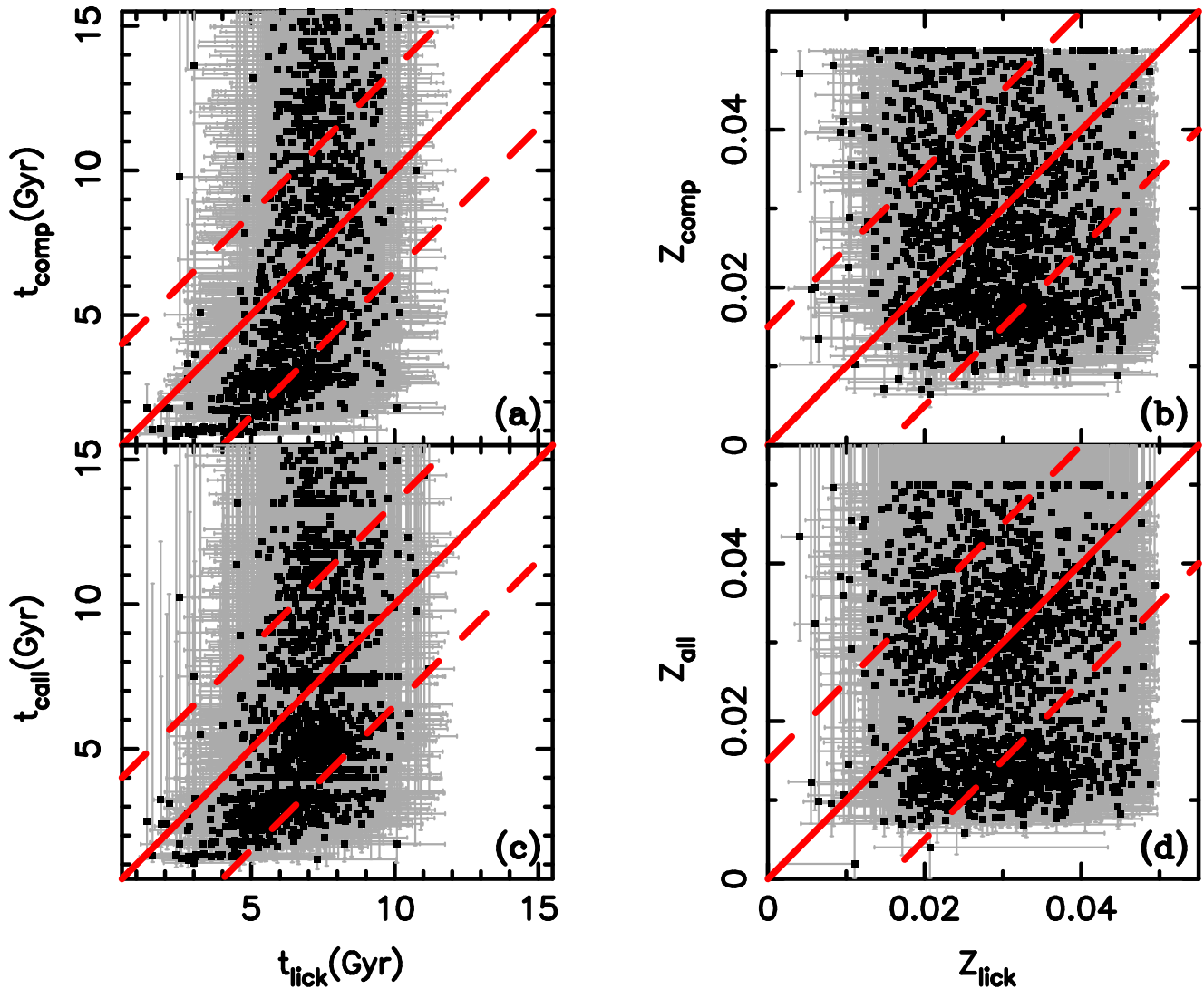}
 \caption{Stellar-population parameters measured using colours
 in conjunction with line indices are plotted versus those measured by a few line indices.
 The suffix ``comp'' represents the results derived from H$\beta$ and $(r-K)$
 while ``all'' for the results measured by two colours in conjunction with two line indices.
 Lines have meanings similar to Fig. 4. The typical uncertainties for stellar ages and metallicities are 4 Gyr and 0.015,
 respectively.}
\end{figure*}

\section{discussions and conclusions}

We investigated the relative metallicity sensitivities of AB system
colours relating to $u$, $B$, $g$, $V$, $r$, $R$, $i$, $I$, $z$,
$J$, $H$, and $K$ bands in the first place. Then we studied the
abilities of colour pairs for constraining luminosity-weighed
stellar ages and metallicities. The results showed that [$(r-K),
(u-R)$] and [$(r-K), (u-r)$] are the best colour pairs for breaking
the stellar age--metallicity degeneracy while colour pairs such as
$(R-K), (u-R)$], [$(I-K), (u-R)$], [$(R-K)], (u-r)$] and [$(i-J),
(u-R)$] can also be used. Colour pairs [$(r-K), (u-R)$] and [$(r-K),
(u-r)$] can measure two stellar-population parameters with small
uncertainties ($\overline{\Delta t} \leq$ 3.89 Gyr,
$\overline{\Delta [Z/{\rm H}]} \leq$ 0.34 dex for typical
uncertainties in colours). However, the age uncertainties for old
populations (Age $\geq$ 4.6 Gyrs) and metal-poor populations ($Z$
$<$ 0.001) are always significantly larger than for young
populations (Age $<$ 4.6 Gyrs) and metal-rich populations ($Z$
$\geq$ 0.001). The reason is that the colours of some old
populations or metal-poor populations are largely indistinguishable
within present typical errors. One can see Fig. 3 for comparison.
However, the metallicity uncertainty of metal-poor populations
(about 0.0024) is much less than that of other populations (about
0.015). In the work, we did not take the uncertainties in
theoretical stellar population models into account. A detailed study
about it please refer to the work of \citet{Yi:2003}. Furthermore,
colours are usually affected by the dust of galaxies, even if
early-type ones. A typical effect of dust in early-type galaxies,
E(B-V)$ \sim$ 0.05, will change the $(r-K)$ and $(u-r)$ colours of
galaxies about 0.12 mag, and then lead to additional uncertainties
in stellar-population parameters (about 2.5 Gyr for age and 0.015
for metallicity). Thus it is more suitable to use colours to measure
the stellar-population parameters of galaxies with poor dust and
gas, e.g., luminous early-type ones (with small colour
uncertainties). Actually, to quantify the age and metallicity errors
induced by a typical dust is in study. When we compare the results
determined by photometry with those determined by Lick indices, we
found that the stellar ages determined by colours are less than
those determined by Lick indices. The difference mainly results from
the effects of young populations in these galaxies. We also found
that the average uncertainties of stellar-population parameters
determined by colours and Lick indices are similar (typically 4 Gyr
for age and 0.015 for metallicity). Therefore, it is actually
difficult to get the absolute values of stellar ages and
metallicities via colours, but we can get some relative values. This
will be useful for some statistical studies of the stellar
populations of galaxies. In addition, our results suggest that it is
better to use colours relating to both $UBVRIJHK$ and $ugriz$ bands
than to use those only relating to one of the two kinds of bands for
estimating the two stellar-population parameters. According to the
results, we can estimate the stellar-population parameters of some
distant galaxies, via the photometry data supplied by, e.g., SDSS
and 2MASS. The possible uncertainties of using various colour pairs
can be estimated. The results presented in the paper can also help
us to choose suitable bands for the observation of stellar
population studies.

We also tried to find some colour pairs that are suitable for
estimating the luminosity-weighted ages and metallicities of some
special stellar populations. The results show that some colour pairs
for estimating the two stellar-population parameters of young (Age
$<$ 4.6 Gyr), old (Age $\geq$ 4.6 Gyr), metal-poor ($Z$ $<$ 0.001),
and metal-rich ($Z$ $\geq$ 0.001) populations are different.
However, [$(r-K), (u-r)$] can be used to estimate the ages and
metallicities of all stellar populations.

Although we took the BC03 model for our work, the results can be
used for other models as most stellar population synthesis models
predict similar colours for the same population. Furthermore, we
suggest to choose a colour pair for estimating the two
stellar-population parameters of galaxies, although a few colours
can give smaller range for the two parameters. The reason is that
galaxies usually contain more than two populations, and via
comparing observational colours with predictions of simple stellar
populations, we can only measure less stellar ages and richer
stellar metallicities for galaxies compared to the dominant stellar
populations (DSPs) of galaxies. For a sample of galaxies, the
effects of young populations on the ages and metallicities of
dominant stellar populations can be roughly corrected and some
reliable estimations for the averages and distributions of two
stellar-population parameters can be obtained
\citep{Li:2007effects}.

Note that the BC03 model is a single stellar population model, which
does not take the binary interactions into account. This is actually
different from the real stellar populations of galaxies. If stars of
a population evolve as binaries rather than single stars, the
colours of the population will be different with those of a single
stellar population. Typically, the $(u-r)$ and $(r-K)$ colours
predicted by binary stellar populations will be bluer about 0.05 mag
than those predicted by single populations. Using binary populations
instead of single populations, the stellar ages and metallicities
will be about 1.14 Gyr older and 0.0093 richer, respectively. A
detailed study about this will be given in the future.

\section*{Acknowledgments}

We greatly acknowledge the anonymous referee for some useful
comments. We also thank Profs. Gang Zhao, Xu Kong, and Dr. Fenghui
Zhang for useful discussions, Dr. Anna Gallazzi for some useful
discussions and her group for the line indices, stellar ages and
metallicities of our sample galaxies, the Sloan Digital Sky Survey
(SDSS), the Two-Micron All-Sky Survey (2MASS) and the NASA/IPAC
Extragalactic Database (NED) for the photometry data of galaxies.
This work is supported by the Chinese National Science Foundation
(Grant Nos 10433030, 10521001), and the Chinese Academy of Science
(No. KJX2-SW-T06).


\bsp

\label{lastpage}

\end{document}